\documentstyle[prl,aps,epsfig,floats]{revtex}
\def\DESepsf(#1 width #2){\epsfxsize=#2 \epsfbox{#1}}
\begin{document}
\pagestyle{empty}                                      
\preprint{
\hbox to \hsize{
\hbox{
            }
\hfill $
\vtop{
 \hbox{ }}$
}
}
\draft
\vfill
\twocolumn[\hsize\textwidth\columnwidth\hsize\csname 
@twocolumnfalse\endcsname
\title{Phenomenological Consequences of Right-handed Down Squark Mixings}
\vfill
\author{Chun-Khiang Chua and Wei-Shu Hou }
\address{
\rm Department of Physics, National Taiwan University,
Taipei, Taiwan 10764, R.O.C.
}

%
%
\vfill
\maketitle
\begin{abstract}
The mixings of $d_R$ quarks,
hidden from view in Standard Model (SM),
are naturally the largest if one has an Abelian flavor symmetry.
With supersymmetry (SUSY) their effects can surface 
via $\tilde d_R$ squark loops. 
Squark and gluino masses are at TeV scale, 
but they can still induce effects comparable to SM 
in $B_d$ (or $B_s$) mixings, 
while $D^0$ mixing could be close to recent hints from data. 
In general, $CP$ phases would be different from SM, 
as may be indicated by recent B Factory data.
Presence of
non-standard soft SUSY breakings with large $\tan\beta$ 
could enhance $b\to d\gamma$ (or $s\gamma$) transitions.
\end{abstract}
\vskip2pc]

\pagestyle{plain}

With the impressive advent of B Factories,
we are entering a new episode of 
flavor and $CP$ violation studies.
Charmless rare $B$ decays hint that 
$\phi_3 \equiv \arg V_{ub}^*$ (PDG phase convention~\cite{PDG})
may be larger than~\cite{Hou} suggested by
a CKM unitarity fit~\cite{Stocchi}.
First results from both B Factories~\cite{sin2phi1} give
smaller $\sin2\phi_1$ (or $\sin2\beta$,
$\phi_1 \equiv \beta\ \equiv \arg V_{td}^*$) values 
than ``expected"~\cite{Stocchi}.
A plausible picture is that $B_d$ or $B_s$ mixings have
additional new physics sources
which invalidate $\Delta m_{B_s}/\Delta m_{B_d}$ constraint,
a notion that can be further probed at B Factories,
or at the Tevatron collider starting next year.
There are also recent hints~\cite{xD} for $D^0$ mixing, which, 
if one has a strong phase difference~\cite{strongphase} 
between $D^0 \to K^+\pi^-$ and $K^-\pi^+$ amplitudes,
can again be taken as hinting at new physics.
In this Letter we study a generic flavor violation scenario
whereby $B_d$ (or $B_s$) and $D^0$ mixings,
as well as $b\to d\gamma$ (or $s\gamma$) transitions, 
could be measurably affected.

New physics in the flavor sector is likely
since little is understood.
For example, fermion masses and mixings
exhibit an intriguing hierarchical pattern
in powers of $\lambda \equiv \vert V_{us}\vert$.
It suggests \cite{horizontal} a possible underlying symmetry, 
the breaking of which gives an expansion in 
$\lambda \sim \langle S\rangle/M$, with
$S$ a scalar field and $M$ a high scale.
If this ``horizontal" (in flavor space) symmetry is Abelian, 
the commuting nature of horizontal charges in general gives
$M_{ij} M_{ji} \sim M_{ii} M_{jj}$ ($i$, $j$ not summed), 
hence \cite{Nir}
\begin{equation}
{M_u\over m_t} \sim \left[
\matrix{\lambda^7 &\lambda^5 &\lambda^3 \cr
         \lambda^6 &\lambda^4 &\lambda^2 \cr 
         \lambda^4 &\lambda^2 &1}
\right],\;
{M_d\over m_b}\sim 
\left[
\matrix{\lambda^4 &\lambda^3 &\lambda^3 \cr
         \lambda^3 &\lambda^2 &\lambda^2 \cr 
         \lambda &1 &1}
\right].
\label{Mq}
\end{equation}
The diagonal elements correspond to quark masses  while
the upper right (lower left) parts are diagonalized by
$U_{qL}$ ($U_{qR}$) rotations.
We note that $M_d^{32}/m_b$ and $M_d^{31}/m_b$ are 
the most prominent off-diagonal elements~\cite{Dudas}. 
Thus, $B_d$ and $B_s$ mixings are naturally susceptible to 
new physics arising from the right-handed down flavor sector.

As a leading candidate for physics beyond SM,
SUSY offers a large toolbox for phenomenology.
For example, squark mixings 
could generate flavor changing neutral currents (FCNC) 
because of strong $\tilde q$-$\tilde g$ coupling. 
It is customary to take squarks as 
almost degenerate at scale $\widetilde m$,
and the squark mixing angle in quark mass basis is
\begin{equation}
\delta^{ij}_{qAB} \cong
[U^\dagger_{qA}\,(\widetilde M^2_q)_{AB}\, U_{qB}]^{ij}
                         /{\widetilde m}^2,
\label{delta}
\end{equation}
where $A,B=L,R$, $i,j$ are generation indices and 
$\widetilde M^2_q$ squark mass matrices. 
As $U_{qL}$ is constrained by the CKM matrix $V$,
mixing angles in $U_{qR}$ are in general larger.
If the breakings of flavor symmetry and SUSY are not closely related, 
then $(\widetilde M^2_q)_{LR}$ 
and $(\widetilde M^2_q)_{RL}=(\widetilde M^2_q)_{LR}^\dagger$ are 
{\it roughly} proportional to respective quark mass matrices,
hence their effects are suppressed by $m_q/\widetilde m$.
From Eq.~(1), one easily gets
${(\widetilde M^2_Q)_{LL}/ \widetilde m^2} \sim~V$, while
\begin{equation}
(\widetilde M^{2}_d)_{RR}
\sim \widetilde m^2 \left[
\matrix{1 &\lambda &\lambda \cr
         \lambda &1 &1 \cr 
         \lambda &1 &1}
\right],
\label{MRR}    
\end{equation}
illustrating that the $RR$ sector could 
contribute significantly to $B_d$ and $B_s$ mixings
via $\delta^{13}_{dRR}\sim\lambda$ and $\delta^{23}_{dRR} \sim 1$.

Defining $x\equiv m_{\tilde g}^2/\widetilde m^2$,
one has
the effective Hamiltonian 
 $H_{\rm eff}=-(\alpha^2_s/216 {\tilde m}^2)
(C_1 {\cal O}_1 + \tilde C_1 \tilde {\cal O}_1
 + C_4 {\cal O}_4 + C_5 {\cal O}_5)$
induced by gluino-squark box diagrams,
where ${\cal O}_1 = \bar d^\alpha_{iL}\gamma_\mu b^\alpha_L\,
           \bar d^\beta_{iL}\gamma^\mu b^\beta_L$,
${\cal O}_{4(5)} = \bar d^\alpha_{iR} b^{\alpha(\beta)}_L\,
           \bar d^\beta_{iL} b^{\beta(\alpha)}_R$,
and~\cite{Gabbiani}
\begin{eqnarray}
C_1 &=&\bigl[24\,x f_6(x)+66 \tilde f_6(x)\bigr]\, 
                (\delta^{i3}_{LL})^2,
\nonumber \\
C_{4(5)} &\cong& \bigl[504(24)\,x f_6(x)-72(+120)\tilde f_6(x)\bigr]\,
                \delta^{i3}_{LL}\delta^{i3}_{RR},
\label{wilson}
\end{eqnarray}
with $\tilde C_1 \tilde {\cal O}_1$ obtained from
$C_1 {\cal O}_1$ by $L\rightarrow R$.
Chargino and neutralino box diagram contributions
are incorporated but numerically unimportant.
Taking into account QCD running~\cite{Bagger},
it is easy to evaluate the $B$ mixing parameter
$M^B_{12} \equiv \vert M^B_{12}\vert \, e^{2i\phi_B}
 = \vert M^{\rm SM}_{12}\vert \, e^{2i\phi_1}
 + \vert M^{\rm SUSY}_{12}\vert \, e^{i\phi_{\rm SUSY}}$.

With $\vert V_{ub}\vert = 0.41\vert\lambda V_{cb}\vert$,
$\phi_3 =65^\circ,\,85^\circ$,
we get $|V_{td}|\times10^{3}=8.0,\,9.2$ and 
$\Delta m_{B_d}^{\rm SM} \sim 0.41,\,0.55\,{\rm ps}^{-1}$, 
respectively, compared with $\Delta m_{B_d}^{\rm exp} 
=0.472\pm0.017\,{\rm ps}^{-1}$~\cite{PDG}.
Allowing $\vert M^{\rm SUSY}_{12}\vert$ to be 
at most of similar size, we find that 
$m_{\tilde q} \sim \widetilde m$, $m_{\tilde g}$ 
cannot be much lighter than 1.5 TeV. 
For illustration we plot, in Fig. 1(a) and (b),
$\Delta m_{B_d} \equiv 2\vert M^B_{12}\vert$ and 
$\sin2\phi_{B_d}$ vs. $\phi \equiv \arg \delta_{dRR}^{13}$, 
respectively, for $\tan\beta =2$ and $|\mu|<m_{\tilde g}$.
We see that $\sin2\phi_{B_d}$ as measured from $B_d \to J/\psi K_S$ 
can range from 0.3 to 1, 
as compared to $\sin2\phi_1 \simeq 0.75$--$0.71$ 
for $\phi_3 = 65^\circ$--$85^\circ$ in~SM.
{\it The low $\sin2\phi_{B_d} \sim 0.3$--$0.4$ possibility 
may be of particular interest} 
when compared to BaBar and Belle central values~\cite{sin2phi1}.
Whether $B_s$ mixing is similarly affected,
it is clear that the CKM unitarity bound from
$\Delta m_{B_s}/\Delta m_{B_d}$ \cite{Stocchi} should be relaxed,
and the potential conflict on $\phi_3$ value with respect to
charmless rare $B$ decays \cite{Hou} may be alleviated.
With $\widetilde m$, $m_{\tilde g} \gtrsim$ TeV 
and $LR$ squark mixings suppressed by quark masses,
there is little impact on penguins, 
and charmless $B$ decays have better access to CKM unitarity phases, 
except being clouded by hadronic uncertainties.

\begin{figure}[t!]
\centerline{
            {\epsfxsize1.65 in \epsffile{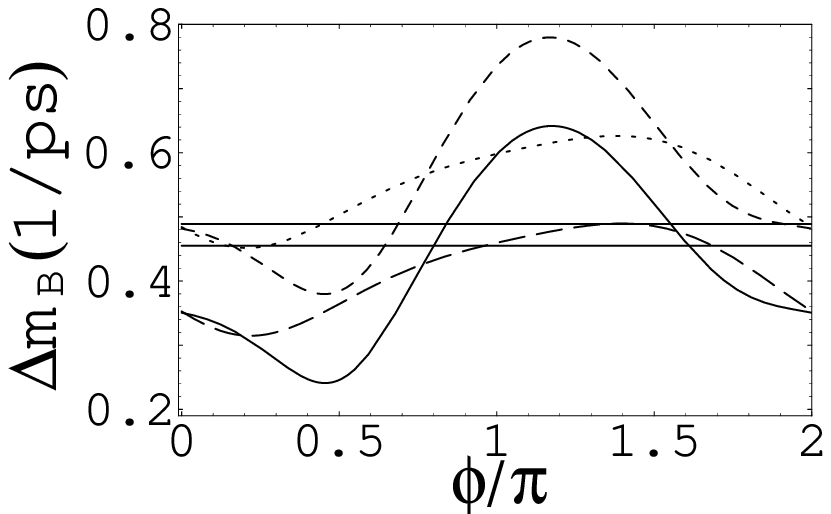}}
            {\epsfxsize1.65 in \epsffile{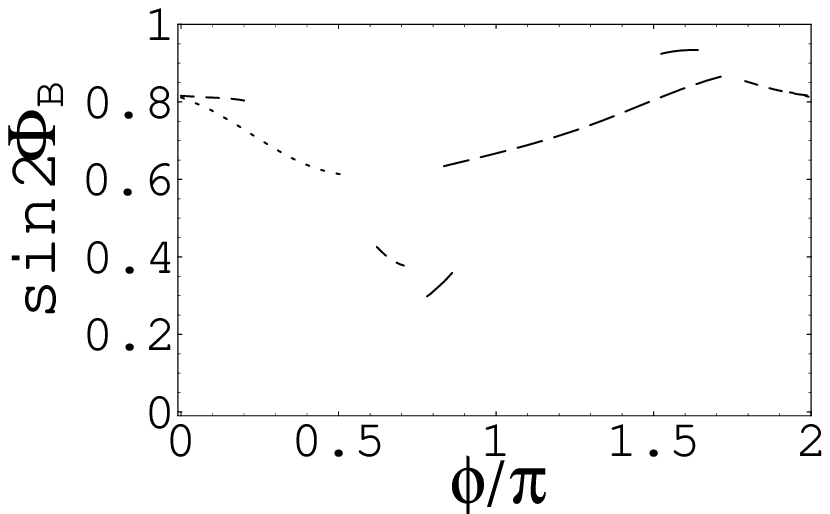}}}
\smallskip
\caption {
(a) $\Delta m_{B_d}$ and (b) $\sin2\phi_{B_d}$
 vs. $\phi \equiv \arg \delta_{dRR}^{13}$, 
including both SM and SUSY effects, 
for squark mass $\widetilde m =$ 1.5 TeV. 
The solid (short-dash), long-dash (dotted) curves 
correspond to $m_{\tilde g} = 1.5$, 3 TeV 
for $\phi_3 = 65^\circ$ ($85^\circ$), respectively. 
The~horizontal lines in (a) indicate the $2\sigma$ experimental range;
theoretical error would allow larger $\sin2\Phi_B$ range than shown.
}
\label{fig:Bd}
\end{figure}

Eqs. (\ref{Mq}) and (\ref{MRR}) are of course too naive.
Since $\Delta m_K$ is much smaller 
than $\Delta m_{B_d}$ while $\varepsilon_K$ is even tighter,
$\delta^{12}_{dLL,RR} \sim \lambda$ is impossible to sustain,
even with $\widetilde m$, $m_{\tilde g} \gtrsim$ TeV.
Through formulas similar to Eq. (\ref{wilson}),
the most severe constraint is on 
$\delta^{12}_{dLL}\delta^{12}_{dRR}$,
which has to be suppressed by approximate ``texture" zeros,
achieved by invoking quark-squark alignment \cite{Nir}.
By using two (or more) singlet fields $S_i$ to 
break the $U(1)\times U(1)$ (or higher) Abelian horizontal symmetry,
and making use of the holomorphy nature of 
the superpotential in SUSY models, one can have $M_d^{12,21}=0$ 
which implies $U^{12}_{dL,R}=0$ (or highly suppressed),
and likewise $(\widetilde M^2_d)_{LL,RR}^{12}$ are also suppressed.
Thus, $\delta^{12}_{dLL,RR}$ can be suppressed
and the kaon mixing constraint satisfied accordingly. 

There is one subtlety arising from our
choice of nonvanishing $M_d^{31}$.
Since $M_d$ is diagonalized by bi-unitary transform,
$M_d^{23}/m_b$, $M_d^{31}/m_b \sim \lambda^2$, $\lambda$
are absorbed by $U^{23}_{dL}$, $U^{13}_{dR} \sim \lambda^2$, $\lambda$,
respectively.
Imposing $M_d^{12}=M_d^{21}=0$,
one finds $U^{12}_{dR} \sim \lambda$ is still needed 
to ensure $M_d^{21} \cong 0$.
From Eq. (\ref{delta}) we see that $\delta^{12}_{dRR} \sim \lambda$
would again be generated, which is not acceptable.
If we now set $M_d^{23}/m_b$ to zero but
retain $M_d^{32}/m_b \sim 1$, 
one finds $U^{23}_{dL} \sim \lambda^2$ is still needed,
again leading to $U^{12}_{dR} \sim \lambda$. 
Thus, to keep 
$(\widetilde M^2_d)_{RR}^{13}/\widetilde m^2 \sim \lambda$,
we need to decouple $s$ flavor from other generations,
i.e. $M_d^{23}$ {\it and} $M_d^{32}$ both set to zero,
and we would have no new physics effects
in $B_s$ mixing and $b\to s\gamma$ decays.
Stringent $\Delta m_K$ and $\varepsilon_K$ constraints
lead to 4 texture zeros in $M_d$.
In the usual approach of quark-squark alignment,
one drops $M_d^{31}$ and $M_d^{32}$ 
(hence $\delta^{13}_{dRR}$ and $\delta^{23}_{dRR}$) 
to satisfy $B_d$ mixing and $b\to s\gamma$ constraints,
allowing for lower $m_{\tilde q}$, $m_{\tilde g}$
that can give collider and other signatures.
In an earlier work we considered 
decoupling the $d$ flavor~\cite{b2sp},
which again has 4 texture zeros.
We will return to a discussion of
this case later.

\begin{figure}[b!]
\centerline{
            {\epsfxsize2.75 in \epsffile{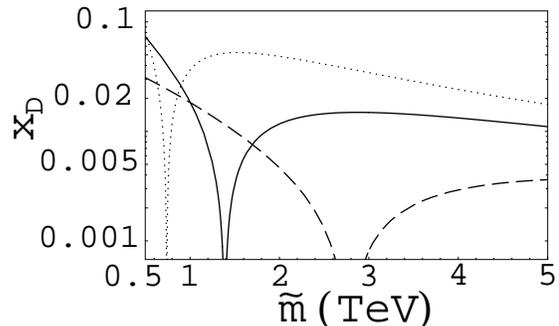}} }
\caption{
$x_D$ vs. $\widetilde m$.
Dotted, solid and dash curves are for 
$m_{\tilde g}= 0.8$, 1.5  and 3 TeV, respectively,
for $\tan\beta = 2$.
}
 \label{fig:Dmix}
\end{figure}

A general consequence of quark-squark alignment~\cite{Nir}
is worthy of note:
$U^{12}_{dL} \simeq 0$ implies
$U^{12}_{uL}\sim \vert V_{cd}\vert =\lambda$,
which would generate $D^0$--$\bar D^0$ mixing 
since $\delta^{12}_{uLL} \sim \lambda$, 
as one can see from Eq. (\ref{delta}).
This is of interest since recent data
hint at~\cite{xD} the possible existence of $D^0$ mixing.
The FOCUS experiment reports a 2.2$\sigma$ deviation of 
the lifetime ratio of $D^0\to K^-K^+$ vs. $K^-\pi^+$ from~1, 
while the CLEO experiment reports a
1.8$\sigma$ effect on $y_D^\prime=y_D\cos\delta_D-x_D\sin\delta_D$
where $x_D \equiv \Delta m_D/\Gamma_D$, 
$y_D \equiv \Delta \Gamma_D/\Gamma_D$,
and $\delta_D$ is the relative strong phase between 
$D^0\to K^+\pi^-$ and $K^-\pi^+$ decay amplitudes.
The two results can be better reconciled 
if $\delta_D \neq 0$~\cite{strongphase}.
While it is certainly premature to conclude that 
one has nonvanishing $x_D$ (which would imply new physics),
what we find intriguing is that
{\it $\delta^{12}_{uLL} \sim \lambda$ with 
$\widetilde m$, $m_{\tilde g} \sim $ TeV
brings $x_D$ right into the ballpark of present sensitivities!}
We illustrate $x_D$ vs. $\widetilde m$ in Fig. 2
for several $m_{\tilde g}$ values $\gtrsim$ TeV.
The zeros reflect a possible cancellation between 
various terms in case the $\delta$'s have common phase.
In practice this is unlikely, since 
the SUSY phases in $\delta^{12}_{uLL,RR}$ are largely unconstrained, 
but it illustrates the adjustability of $x_D$. 
However, one has an explicit example where 
detectable $D^0$ mixing would likely \cite{Wolf_D} 
carry a $CP$ violating phase.

Having satisfied $\Delta m_K$, $\varepsilon_K$ by construction,
one can still have interesting and measurable effects in
$B_d$ and $D^0$ mixings 
even if SUSY breaking is at TeV scale.
This is because of large $\tilde d_R$-$\tilde b_R$
and $\tilde u_L$-$\tilde c_L$ mixings
which follow from Abelian horizontal charges and low energy constraints.
Unfortunately,
the SUSY scale is so high such that
there is practically no impact on penguins such as
$\varepsilon^\prime/\varepsilon$, $b\to s\gamma$ and $b\to d\gamma$.
One also has the depressing situation that
the squarks and gluino cannot be produced at the Tevatron or LHC. 
While we cannot change the latter,
we find that radiative penguins can pick up 
some exotic effects in SUSY breaking,
such as the presence of non-standard soft breaking $C$-terms.

We are 
interested in 
{\it bona fide} evidence for physics beyond SM.
One such effect is mixing dependent $CP$ violation in
$b\to s\gamma$ or $d\gamma$ transitions\cite{Atwood}.
The effective Hamiltonian 
is given by 
$
H_{\rm eff.} \propto 
m_b\,\bar d (C_7\,R+C_7^\prime\,L) \sigma_{\mu \nu }F^{\mu \nu }b.
$
Mixing dependent $CP$ violation in $B^0 \to M^0\gamma$ decay
is rather analogous to the golden $J/\psi K_S$ mode, i.e.
\begin{equation}
a_{M^0\gamma}=\xi \sin 2 {\vartheta}\, 
                  \sin[2\phi_B-\phi(C_7)-\phi(C^\prime_7)]\,
                  \sin \Delta m\,t
\end{equation}
where $CP(M^0)=\xi M^0$,
$\phi(C^{(\prime)}_7)$ is the phase of $C^{(\prime)}_7$, and 
$
\sin 2 {\vartheta}\equiv 
2\,\vert C_7 C_7^\prime\vert\,/\,(\vert C_7\vert^2+\vert C_7^\prime\vert^2).
$
One clearly needs both $C_7$ and $C^\prime_7$ 
for the interference to occur.
In SM, however, $C_7^\prime/C_7 \propto m_{d,s}/m_b$ 
and is hence negligible.
Thus, $a_{M^0\gamma}$ is sensitive to new physics \cite{Atwood}.

At $\widetilde m$ scale the SUSY contribution is
\begin{eqnarray}
C_{7,\tilde g}^\prime &\propto& g_s^2 Q_d 
      [\delta^{13}_{dRR} \, g_2(x)
     - \delta^{13}_{dRL}{m_{\tilde g}\over m_b}  g_4(x)\, 
       ]/G_F\widetilde m^2,
\label{c7pgluino}
\end{eqnarray}
where $g_l(x) = -{d/dx} (x\,F_l(x))$ 
with $F_l(x)$ taken from~\cite{Bertolini}.
Exchanging $L \leftrightarrow R$ gives the correction to $C_7$,
and QCD running can be taken from~\cite{c7c8mix}.
The chiral or $RL$ enhancement 
in Eq. (\ref{c7pgluino}) is apparent, 
noted already in our study~\cite{b2sp} 
of large $\tilde s$-$\tilde b$ mixings.
It has also been invoked to generate $\varepsilon^\prime/\varepsilon$
via an analogous $\delta^{12}_{LR}$ term~\cite{Masiero}
under a horizontal U(2) (hence non-Abelian) symmetry model.

We see that large $\delta^{13}_{dRR}$ 
or $\delta^{13}_{dRL}\, m_{\tilde g}/m_b$ are needed,
which is precisely our case with Abelian horizontal symmetry. 
The $RL$ enhancement factor $m_{\tilde g}/m_b$ can compensate for 
quark mass suppression in $(\widetilde M_d^2)_{RL}$.
Unfortunately, the high SUSY scale
leads to too severe a suppression in $1/G_F\widetilde m^2$.
We find, however, that large $a_{M^0\gamma}$ is still possible 
when considering certain
{\it non-standard soft breaking terms} \cite{Hall,diazcruz}.
Allowing for a non-standard $C \langle H^*_u\rangle 
 Y^{\prime\prime} \tilde D_L \tilde D_R^*$ \cite{Cterm}
besides the standard $A_d \langle H_d\rangle
 Y^\prime \tilde D_L \tilde D_R^*$,
it is natural that $A_d \sim C \sim \widetilde m$, hence
$(\widetilde M^2_d)^{ij}_{LR} \sim \widetilde m\,M^{ij}_d\,\tan \beta$,
and one gains a $\tan\beta \equiv
 \vert\langle H^*_u\rangle/\langle H_d\rangle\vert$
enhancement factor, while
$(\widetilde M^2_u)^{ij}_{LR}\sim \widetilde m\,M^{ij}_u$ is unaffected.
Some zeros in $(\widetilde M^2_q)_{LR}$ may be lifted 
since these $C$-terms are no longer holomorphic, 
but they are still suppressed.
Arising from the superpotential, $M_q$, and hence $U_{qL,R}$, 
is unchanged, so the result for $D^0$ mixing remains unchanged. 
The $\delta^{12}_{dRL}$ contribution to kaon mixing remain 
protected by the smallness of $M^{21}_d/{\widetilde m}$.
Likewise, for $B_d$ mixing,
$\tan\beta$ enhancement of $\delta_{dLR,RL}^{13}$
is insufficient to overcome $m_q/\widetilde m$ suppression
and $\delta^{13}_{dRR}$ still dominates.

We illustrate in Fig. 3(a) and (b) the ratio
${\rm Br}(B\to X_{d}\gamma)/{\rm Br}(B\to X_{d}\gamma)_{\rm SM}$
and $\sin 2\vartheta$ of Eq. (5) with respect to $\widetilde m$,
for $m_{\tilde g}= 1.5$ TeV and $\tan\beta=50,\,20$ and 2 
(this also illustrates the case of standard $A$-term only).
The rate enhancement over SM can reach 
a factor of 5, 1.8 and few \%, respectively,
{\it hence Br($B\to\rho\gamma$) could reach $10^{-5}$ level,
while $\sin 2\vartheta$ can easily reach maximum 
for large $\tan\beta$}.
We note that current limits on $B\to\rho\gamma$
are beginning~\cite{Nakao} to probe such levels, 
as the B Factories have demonstrated their 
$K/\pi$ (hence $K^*/\rho$) separation capabilities, 
and data is accumulating fast.
There is a further advantage in studying 
mixing dependent $CP$ asymmetries 
$a_{\rho^0\gamma}$, $a_{\omega\gamma}$: 
$\rho^0,\ \omega \to \pi^+\pi^-(\pi^0)$ gives vertex information
but {\it not} in $K^{*0}\to K^0\pi^0$ case,
and one would have to go to modes such as 
$B^0\to K^{0}_1\gamma$, where one is penalized by
$K_1^0 \to \rho^0 K^0$ branching ratios.

\begin{figure}[t!]
\centerline{{\epsfxsize1.65 in \epsffile{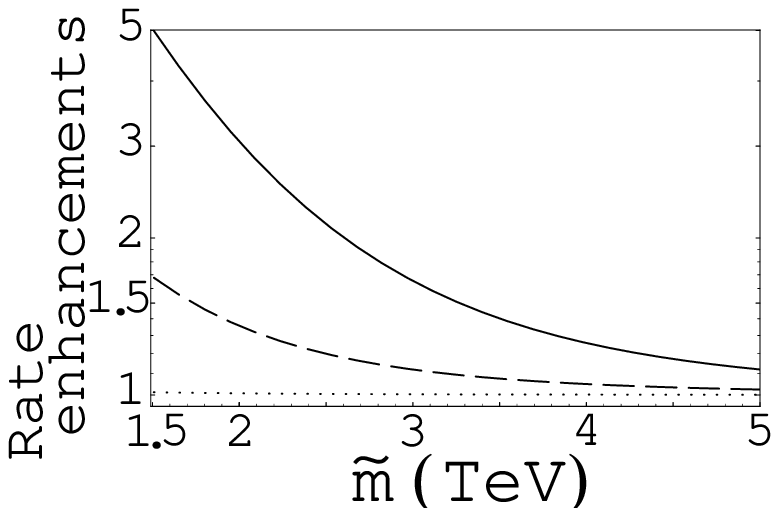}}
            {\epsfxsize1.60 in \epsffile{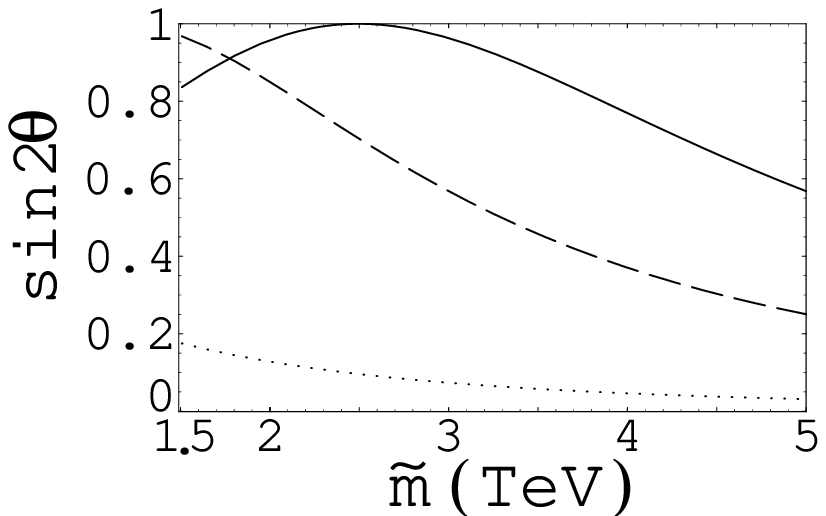}}}
\smallskip
\caption{
(a) ${\rm Br}(B\to X_{d}\gamma)/{\rm Br}(B\to X_{d}\gamma)_{\rm SM}$
and (b)~$\sin 2\vartheta$ vs. $\widetilde m$
for $ m_{\tilde g}= 1.5$ TeV.
The solid, dash and dotted curves 
correspond to $\tan\beta=50,\,20$ and 2, 
respectively.
}
 \label{fig:radiative}
\end{figure}

Our exotic ``non-standard $C$-term",
together with large $d_R$-$b_R$ mixing (Eqs. (1) and (3))
and large $\tan\beta$ (motivated by large $m_t/m_b$ ratio),
give an existence proof for possible prominence of
$B\to\rho^0\gamma$, $\omega\gamma$ modes,
both in rate and mixing dependent $CP$ asymmetries.
They should be priority search and study items for the coming years.
Taking a more liberal point of view,
in principle one can have 
flavor violating soft SUSY breaking terms~\cite{b2sp}
which could easily generate $\delta_{dLR,RL}^{13}$ 
without resorting to non-standard $C$-terms.
Thus, $b\to d\gamma$ search provides a 
more general probe of flavor violation in SUSY.

We now entertain the case of decoupling $d$ flavor
but keeping $M_d^{32} \sim 1$, 
which was studied in \cite{b2sp} from a different perspective.
From consideration of $\Delta m_K$, $\varepsilon_K$ and $D^0$ mixing, 
the squarks and gluinos are again at TeV scale.
But since $\tilde s_R$-$\tilde b_R$ mixing is large,
there may be one squark that is lighter than the rest.
Clearly, $B_s$ mixing can be easily enhanced
hence the CKM unitarity constraint through
$\Delta m_{B_s}/\Delta m_{B_d}$ again should be relaxed.
Thus, a lower $\sin2\phi_{B_d}$ than predicted by CKM fit is possible,
while the $CP$ phase of $B_s$ mixing is likely nonvanishing,
testable at the Tevatron soon.
The model, however, is constrained by $b\to s\gamma$.
Since $\delta_{dRL}^{13}/\delta_{dRL}^{23} \sim \lambda
\sim V_{td}/V_{ts}$ in SM,
we can take Fig. 3 as a rough estimate for $b\to s\gamma$
in present case.
Allowing for 20\% rate uncertainty for the measured
Br$(B\to X_s\gamma)=(3.15\pm0.54)\times10^{-4}$,
we see that for heavier $m_{\tilde q} \sim 3$ TeV case,
$\tan\beta$ up to 20 is allowed,
and $\sin 2\vartheta$ can go up to $\sim 0.5$,
hence $a_{M^0\gamma}$ study should also be pursued.
For lighter $m_{\tilde g}$ such as 1.5 TeV,
the enhancement factor for large $\tan\beta$ starts to
break the good agreement between SM and experiment,
hence it seems one cannot have both large $\tan\beta$ and
$\tilde m$, $m_{\tilde g}$ too light (TeV or less).
This should apply to the lightest squark in present case.
One may alternatively say that, 
despite the stringent constraint of $b\to s\gamma$ rate,
one can still have nontrivial mixing dependent $CP$
asymmetries.

We give explicit charge assignments for the main case studied.
Two $S_i$ fields are used to break the horizontal symmetry,
${\langle S_1\rangle/M} \sim {\langle S_2\rangle/M} \sim
{\tilde\lambda}^{0.5}$, where $\tilde \lambda=0.18$ 
(to fit the smallness of $V_{ub}$ better).
The horizontal charges are $(-1,0)$ and $(0,-1)$ for $S_1$ and $S_2$, 
and
\[
\begin{array}{lll}
Q_1:\,(8,-2), & Q_2:\,(1,3), & Q_3:\,(2,-2), \\
\bar d_1:\,(-2,10\,[5]), \ & \bar d_2:\,(9\,[4],-3), \ & 
\bar d_3:\,(-2,8\,[3]), \\
\bar u_1:\,(-3,11), & \bar u_2:\,(0,3), & \bar u_3:\,(-1,2),
\end{array}
\]
for small $\tan\beta$ [$\tan\beta \sim 50$].
With these assignments,
$M_u^{21}$, $M_u^{31}$ and
$M_d^{21,12}$, $M_d^{23,32}$ vanish,
and corresponding elements in
$(\widetilde M^2_d)^{ij}_{RR}/\widetilde m^2$
become $\sim \tilde \lambda^{12\,[7]}$ or $\tilde \lambda^{11\,[6]}$.
The non-standard $C$-term 
restores the vanishing elements of 
$(\widetilde M^2_d)^{ij}_{RR} \sim \widetilde m\, M_q^{ij}$
to some power, but they do not affect quark mass.
Additional rotations may arise from 
the K\"ahler potential,
which may lift the zeros to the so-called filled zeros \cite{Dudas}.
Their effect is small in this model.
Detailed discussions will be given elsewhere.

Some remarks are in order.
First, 
our numerics are only illustrative,
since the $\delta$'s can not be specified fully.
Second, 
because of stringent $\Delta m_K$ and $\varepsilon_K$ constraints,
$(\varepsilon^\prime/\varepsilon)^{\rm SUSY}$ in this model is 
always very small.
Third, 
the neutron edm is well protected by
$m_d/\widetilde m$ in the generic picture.
However, with non-standard $C$-terms, 
one may need to restrict the phase of $M_d^{11}$
 (and $(\widetilde M^2_d)_{LR}^{11}$ induced by $\mu$)
to 0.1 when $\tan\beta$ is very large (such as 50).
Four, direct $CP$ asymmetries in $b\to d\gamma$ would
get diluted rather than enhanced by SUSY,
especially if $a_{\rho^0\gamma}$, $a_{\omega^0\gamma}$ are large.
Five, for large $\tan\beta$,
the neutralino box diagram contribution to neutral meson mixings
become important, especially if one takes
a GUT motivated relation for gaugino masses, 
which also holds true in gauge-mediated SUSY breaking models.
However, the qualitative features of Figs. 1 and 2 remain the same.
Six, chargino loops involving light stop or chargino may 
give rise to very significant effects~\cite{Pokorski}
for large $\tan\beta$ that may require fine tuning.
To avoid this, $\vert\mu\vert \sim$ TeV scale is needed.
As for stop, 
we have followed Ref.~\cite{Nir} with the tacit assumption 
that flavor and SUSY scale are not too far apart,
hence the stop does not become too light by 
large accumulative renormalization group running
and hence still satisfy Eq. (2).
Finally,
with high $m_{\tilde q}$ and $m_{\tilde g}$ scale 
but no $\tan\beta$ enhancement, 
SUSY induced radiative $c\to u\gamma$ is 
smaller than two-loop SM correction,
while for $t\to c\gamma$ it enhances SM result of $\sim 10^{-13}$
by three orders.

In summary,
$d_R$ quark mixings are naturally the largest in
Abelian horizontal models.
Such flavor (and $CP$) violation effects
can be brought to light by $\tilde d_{jR}$ squark loops.
Stringent $K^0$ mixing and $\varepsilon_K$ constraints
require setting $M_d^{12} = M_d^{21} = 0$.
If we choose to retain $M_d^{31}/m_b \sim \lambda$,
the $s$ flavor has to be decoupled,
and interestingly one does not have to 
face $b\to s\gamma$ constraint.
Squarks and gluinos have to be at TeV scale,
but they can shift $\sin2\phi_1$ to the 
low value reported by B Factories.
Quark-squark alignment induces $\tilde u_L$-$\tilde c_L$ mixing 
that can give $D^0$ mixing close to 
recent hints from CLEO and FOCUS.
Penguin related phenomena are in general untouched,
but non-standard soft SUSY breaking $C$-terms could,
through large $\tan\beta$ enhancement,
bring $B\to \rho\gamma$, $\omega\gamma$ rates to $10^{-5}$ level,
while mixing dependent $CP$ asymmetries could be maximal.
Similar effects can be induced by generic flavor violating
soft SUSY breaking terms.
If one keeps $M_d^{32}/m_b \sim 1$,
then $d$ flavor has to be decoupled.
$B_s$ mixing can be greatly enhanced with likely
nonvanishing $CP$ phase. One again could have
measurable $D^0$ mixing, while the more constrained
$b\to s\gamma$ transition could have mixing dependent $CP$
asymmetries of order 50\%.
The new physics phenomenology outlined here 
can be tested at B Factories
and the Tevatron in the next few years.

\vskip 0.3cm
\noindent{\bf Acknowledgement}.\ \
This work is supported in part by
the National Science Council of R.O.C.
under Grants NSC-89-2112-M-002-036
and NSC-89-2811-M-002-039.


\begin{thebibliography}{99}

\bibitem{PDG}  
D.E. Groom {\it et al}, 
Eur. Phys. J. C {\bf 15}, 1 (2000).

\bibitem{Hou} X.G. He, W.S. Hou, K.C. Yang, 
Phys. Rev. Lett. {\bf 83}, 1100 (1999);
W.S. Hou, J.G. Smith, F. Wurthwein, hep-ex/9910014. 

\bibitem{Stocchi} F. Parodi, P. Roudeau, A. Stocchi,
Nuovo Cim. A {\bf 112}, 833 (1999);
F. Caravaglios {\it et~al.}, hep-ph/0002171.

\bibitem{sin2phi1} Plenary talks given by
D. Hitlin 
and H. Aihara 
at ICHEP2000, 
July 27 -- August 2, 2000, Osaka, Japan.

\bibitem{xD} 
R. Godang {\it et al.},
Phys. Rev. Lett. {\bf 84}, 5038 (2000);
J.M.~Link {\it et al.},
Phys. Lett. B {\bf 488}, 218 (2000). 

\bibitem{strongphase} A.F. Falk, Y. Nir, A.A. Petrov,
JHEP {\bf9912}, 019 (1999),
S. Bergmann {\it et al.}, Phys. Lett. B {\bf 486}, 418 (2000).



\bibitem{horizontal}
See, e.g. 
C.D. Froggatt, H.B. Nielsen, Nucl. Phys. B {\bf 147}, 277 (1979).

\bibitem{Nir}
Y. Nir, N. Seiberg, Phys. Lett. B {\bf 309}, 337 (1993);
M.~Leurer, Y. Nir, N. Seiberg, Nucl. Phys. B {\bf 420}, 468 (1994).

\bibitem{Dudas}
E. Dudas, S. Pokorski and C.A. Savoy,
Phys. Lett. B {\bf 356}, 45 (1995).
More than half of the examples have large $M^{3j}_d/m_b$,
and our discussion would follow through.

\bibitem{Gabbiani} F. Gabbiani {et al.}, 
Nucl. Phys. B {\bf 477}, 321 (1996).

\bibitem{Bagger} J.A. Bagger, K.T. Matchev, R.J. Zhang, 
                 Phys. Lett. B~{\bf 412}, 77 (1997).

\bibitem{b2sp} C.K. Chua, X.G. He, W.S. Hou, 
Phys. Rev. D {\bf60}, 014003 (1999). 

\bibitem{Wolf_D} L. Wolfenstein,
Phys. Rev. Lett. {\bf 75}, 2460 (1995).

\bibitem{Atwood}  D. Atwood, M. Gronau, A. Soni, 
Phys. Rev. Lett. {\bf 79}, 185 (1997).

\bibitem{Bertolini} S. Bertolini {\it et al.}, 
Nucl. Phys. B {\bf 353}, 591 (1991).

\bibitem{c7c8mix} A.J. Buras {\it et al.}, 
Nucl. Phys. {\bf B424}, 374 (1994).

\bibitem{Masiero} A. Masiero, H. Murayama, 
Phys. Rev. Lett. {\bf 83}, 907 (1999).


\bibitem{Hall} L.J. Hall, L. Randall, 
Phys. Rev. Lett. {\bf 65}, 2939 (1990).

\bibitem{diazcruz} J.L. Diaz-Cruz, hep-ph/9906330.

\bibitem{Cterm} We do not give horizontal charges to
$H_u^*$ and $H_d$ fields, 
hence assume $Y^{\prime\prime} \sim Y^\prime$,
i.e. having same flavor pattern,
and explore simple $\tan\beta$ enhancement 
of $(\widetilde M^2_d)_{LR}$ only.

\bibitem{Nakao} M. Nakao (Belle Collaboration),
talk presented at ICHEP2000, July 27 -- August 2, 2000, Osaka, Japan.


\bibitem{Pokorski} See, e.g. P.H. Chankowski, S. Pokorski,
in {\it Perspectives on Supersymmetry}, ed. G.L. Kane 
(World Scientific).

\end{thebibliography}
\end{document}